\documentclass[11pt,preprintnumbers,aps,amssymb,nofootinbib,amsmath,superscriptaddress,notitlepage]{revtex4-1}
\usepackage{epsfig,epsf}
\usepackage{bm} % puts greek and math symbols in boldface using \bm
\usepackage{color} % {\color{red} ... }
\usepackage{slashed}
\usepackage{relsize}	%rescalable math 
\usepackage{soul} %spacing out (letterspacing), underlining, striking out, etc.
\usepackage{hyperref}
\newcommand{\beq}{\begin{equation}}
\newcommand{\beql}[1]{\begin{equation}\label{#1}}
\newcommand{\eeq}{\end{equation}}
\def\bal#1\gal{\begin{align}#1\end{align}}
\newcommand{\ball}[1]{\bal\label{#1}}
\newcommand{\eq}[1]{(\ref{#1})}
\newcommand{\fig}[1]{Fig.~\ref{#1}}
\renewcommand{\sec}[1]{Sec.~\ref{#1}}
\renewcommand{\b}[1]{{\bm #1}} 
\newcommand{\unit}[1]{\hat {{\bm #1}}} % unit vector

\begin{document}

\title{Magnetic field in expanding quark-gluon plasma}

\author{Evan Stewart}
\author{Kirill Tuchin}

\affiliation{Department of Physics and Astronomy, Iowa State University, Ames, Iowa, 50011, USA}

\date{\today}


\begin{abstract}

Intense electromagnetic fields are created in the quark-gluon plasma by the external ultra-relativistic valence charges. The time-evolution and the strength of this field are strongly affected by the electrical conductivity of the plasma. Yet, it has recently been observed that the effect of the magnetic field on the plasma flow is small. We compute the effect of plasma flow on magnetic field and demonstrate that it is less than 10\%. These observations indicate that the plasma hydrodynamics  and the dynamics of electromagnetic field decouple. Thus, it is a very good approximation, on the one hand, to study QGP in the  background electromagnetic field generated by external sources and, on the other hand,  to investigate the dynamics of magnetic field in the background plasma. We also argue that the wake induced by the magnetic field in plasma is negligible.  

\end{abstract}

\maketitle

%%%%%%%%%%%%%%%%%%%%%%%%%%%%%%%%%%%%%%%%
\section{Introduction}\label{sec:intr}

The main goal of the relativistic heavy-ion collisions program is to produce and study the quark-gluon plasma (QGP). Along with the plasma, the relativistic heavy-ion  collisions produce intense electromagnetic fields that modify its properties. In order to infer the plasma properties from the experimental data one needs to quantify the effect of electromagnetic fields on the QGP dynamics. In principle, this can be accomplished by solving the relativistic magneto-hydrodynamic (MHD) equations. The electromagnetic field affects both the ideal plasma flow and the transport coefficients, while the electric currents in plasma affect the electromagnetic field. 
Since the QGP dynamics is determined mostly by the strong interactions, one may start by treating  the electromagnetic interactions as a small perturbation. This approximation amounts to decoupling, to a certain extent,  of the dynamics of the electromagnetic field and the plasma. 

The MHD of ideal QGP in the background electromagnetic field was studied in \cite{Pu:2016ayh,Roy:2015kma,Pu:2016bxy,Roy:2015coa,Roy:2017yvg,Inghirami:2016iru,Mohapatra:2011ku,Das:2017qfi,Greif:2017irh,Tuchin:2011jw}. It has been recently argued in \cite{Roy:2017yvg} that the effect of the electromagnetic field on QGP is small for realistic fields justifying the decoupling  approximation. Still, before making a final conclusion that the plasma flow is decoupled from the electromagnetic field, one needs  to verify that the kinetic coefficients do not strongly depend on the field. In particular, significant enhancement of the viscous stress  may invalidate the ideal fluidity assumption. Despite the recent progress in calculating the transport coefficients \cite{Ding:2010ga,Aarts:2007wj,Amato:2013oja,Cassing:2013iz,Hattori:2017qih,Yin:2013kya,Li:2017tgi,Hattori:2016idp,Hattori:2016lqx,Fukushima:2015wck,Li:2016bbh,Hattori:2016cnt,Nam:2013fpa,Agasian:2011st,Chernodub:2009rt,Li:2017jwv}, their values at the temperatures of phenomenological interest are not yet certain.

Assuming perfect decoupling, i.e.\ that QGP does not affect the electromagnetic field at all, the electromagnetic field  was computed in \cite{Kharzeev:2007jp,Skokov:2009qp,Bzdak:2011yy,Voronyuk:2011jd,Ou:2011fm,Deng:2012pc,Bloczynski:2012en}  using the hadron transport models. However, it was argued in \cite{Tuchin:2013ie,Tuchin:2013apa} that this approximation is adequate only at the earliest times after the plasma formation. At later times the plasma response plays the crucial role. Owing to its finite electrical conductivity it significantly enhances the electromagnetic field \cite{Tuchin:2010vs,Tuchin:2013apa,Tuchin:2013ie,Zakharov:2014dia,Tuchin:2015oka}. Thus far all calculations of the electromagnetic field assumed stationary plasma. The main  goal of this paper is to compute the contribution of the plasma expansion to the magnetic field. We will argue that this contribution is on the order of a few per cent and thus can be safely neglected. Along the way, we will clarify a number of important points that were not sufficiently addressed in the previous publications.

\begin{figure}[ht]
      \includegraphics[height=5cm]{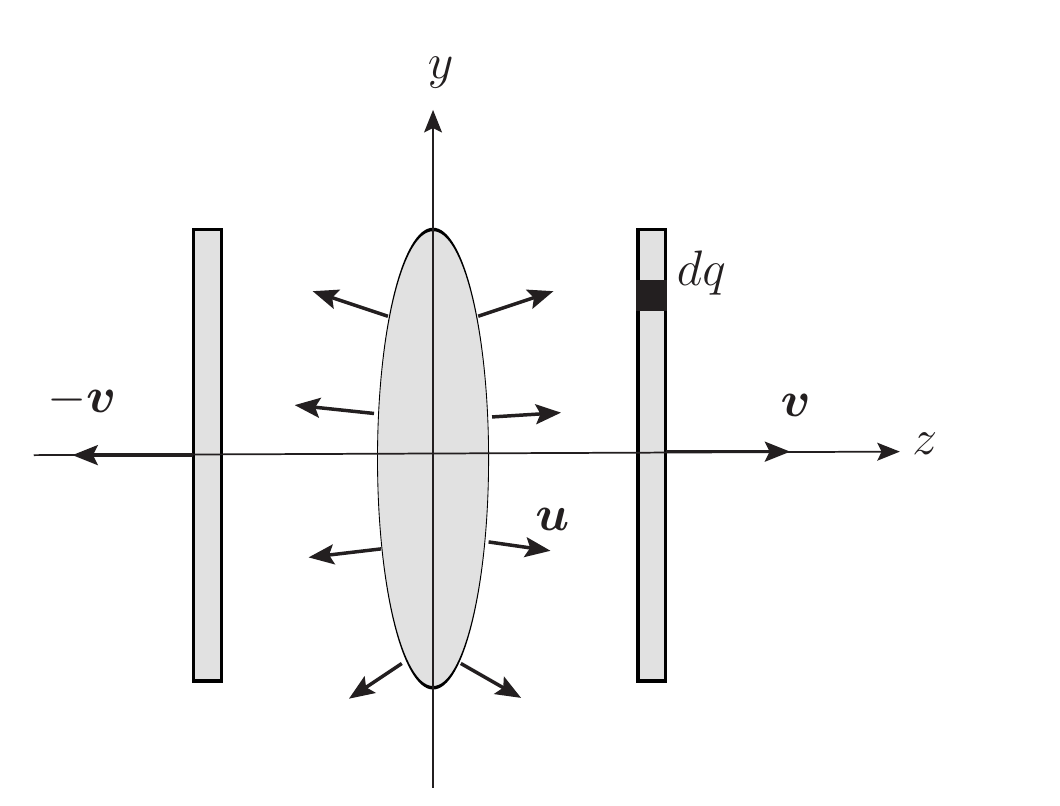} 
  \caption{The geometry of the heavy-ion collisions. Ion remnants move with velocity $\pm \b v$. The plasma's velocity is $\b u$.   We emphasize that the valence electric charges $dq$ are external to the plasma. The geometry in the $xy$ plane is shown in \fig{geom-2}. }
\label{geom-1}
\end{figure}

The spacetime picture of a heavy-ion collision is shown in \fig{geom-1} and \fig{geom-2}. In \fig{geom-1}, which is nearly identical to the one found in the classical Bjorken's paper \cite{Bjorken:1982qr}, we emphasize that the valence quarks, which are sources of the electromagnetic field, are external to the plasma. In fact, a small fraction of valence quarks can be found inside the QGP, which is known as the baryon stopping. However, the transfer of the valence quarks across the wide rapidity interval is strongly suppressed \cite{Kharzeev:1996sq,Itakura:2003jp}. Their contribution to the total field was estimated in \cite{Kharzeev:2007jp} and turns out to be completely negligible at relativistic energies. In view of this observations we neglect the baryon stopping, assuming that all valence quarks travel along the straight lines. Furthermore, for our arguments in this paper it is sufficient to approximate the valence electric charges as classical point particles. In a more  comprehensive treatment one has to replace the classical sources by the quantum distributions 
\cite{Holliday:2016lbx,Peroutka:2017esw}. 

In this paper we regard the QGP as a homogeneous plasma expanding according to the blast wave model \cite{Siemens:1978pb,Teaney:2000cw,Kolb:2000fha} and having the electrical conductivity $\sigma$.  We are going to neglect its mild time dependence \cite{Tuchin:2013ie}  and treat it as a constant \footnote{Actually, even a mild time dependence of $\sigma$ may be phenomenologically significant \cite{Tuchin:2015oka}.}. Recently, there has been a lively discussion of possible effects of the chiral anomaly  \cite{Kharzeev:2013ffa,Huang:2015oca,Kharzeev:2015znc} on the QGP dynamics in general and its electrodynamics in particular \cite{Tuchin:2014iua,Manuel:2015zpa,Li:2016tel,Tuchin:2016qww,Hirono:2016jps,Hirono:2015rla,Xia:2016any,Qiu:2016hzd}. In this paper we adopt a conservative view and disregard these effects until they are firmly established. 

The paper is organized as follows. In \sec{sec:a} we write down the basic equations that determine the electromagnetic fields in QGP.  We derive the retarded Green's function of the electromagnetic field in the electrically conducting medium and show that it is a sum of two terms: the pulse and the wake. 
The wake field is usually neglected in calculations. We prove that this is a good approximation. Indeed, at energies $\gamma=100$ in a plasma with electrical conductivity  $\sigma = 5.8$~MeV \cite{Amato:2013oja,Ding:2010ga}, the wake term is small until $t\sim 100$~fm/$c$  and thus can be neglected in phenomenological calculations. This is discussed in \sec{sec:c} in the stationary plasma limit. The main result of \sec{sec:c}  is Eq.~\eq{a51} which gives the analytical expression for the magnetic field of a point external charge in conducting medium. It agrees with the previous result derived by one of us \cite{Tuchin:2013apa}, but has an advantage of being expressed in terms of the elementary functions. Expanding plasma is considered in \sec{sec:e} were we treat the magnetic part of the Lorentz force perturbatively and derive the solution for the magnetic field. We summarize the results and discuss the prospects in \sec{sec:s}.

%%%%%%
%%%%%

%%%%%%%%%%%%%%%%%%%%%%%%%%%%%%%%%%%%%%%%
\section{Maxwell equations in expanding plasma}\label{sec:a}

 An electromagnetic field in flowing conducting medium satisfies the equations 
\begin{subequations}
\begin{align}
\b \nabla\times \b B &= \partial_t \b E+ \sigma (\b E+\b u\times \b B)+\b j\,,\label{a10}\\
\b \nabla\cdot \b E&= \rho\,, \label{a11}\\
\b \nabla\cdot \b B&= 0\,, \label{a13}\\
\b \nabla \times \b E&= -\partial_t \b B\,, \label{a14}
\end{align}
\end{subequations}
where $\b u$ is the fluid velocity, $\sigma$ is electrical conductivity and $j^\mu = (\rho, \b j)$ is the external current created by the valence charges as shown in \fig{geom-1} . Replacing the fields with the potentials as usual 
\bal
\b E= -\b \nabla \varphi- \partial_t \b A\,, \qquad \b B= \b \nabla\times \b A
\gal
 and using the gauge condition
\ball{a18}
\partial_t \varphi+ \b \nabla \cdot \b A+\sigma\varphi=0\,
\gal
we arrive at the equations
\begin{subequations}
\bal
&-\nabla^2 \varphi+\partial_t^2\varphi + \sigma\partial_t \varphi=\rho\,,\label{a20}\\
&-\nabla^2 \b A+\partial_t^2\b A + \sigma\partial_t \b A-\sigma \b u \times (\b \nabla\times \b A)=\b j\,,\label{a21}
\gal
\end{subequations}

We consider a point charge $e$ moving in the positive $z$ direction with constant velocity $v$: 
\ball{a23}
\b j = ev \unit z\delta(\b b)\delta(z-vt)\,, \quad \rho=0\,.
\gal
In the experimentally interesting region of small $z$'s, $|\b u|\ll 1$. This allows us to treat the corresponding term in \eq{a21} as a perturbation. Thus, writing $\b A= \b A^{(0)}+ \b A^{(1)}$ we obtain two equations
\begin{subequations}
\bal
&-\nabla^2 \b A^{(0)}+\partial_t^2\b A^{(0)} + \sigma\partial_t \b A^{(0)}=\b j\,,\label{a25}\\
&-\nabla^2 \b A^{(1)}+\partial_t^2\b A^{(1)} + \sigma\partial_t \b A^{(1)}= \sigma\b u \times \b B^{(0)}\,.\label{a26}
\gal
\end{subequations}
The first of these equations describes the field created by the external currents in the stationary plasma, whereas the second one takes expansion of plasma into account.  

%%%%%%%%%%%%%%%%%%%
%\section{The Green's functions}\label{sec:b}

To find the particular solutions to these equations we introduce the retarded Green's function $G(\b r, t| \b r', t')$ that obeys the equation
\ball{a28}
&-\nabla^2  G+\partial_t^2 G+ \sigma\partial_t G=\delta(t-t')\delta(\b r-\b r')\,.
\gal
We note that  the function $\mathcal{G}$ defined as
\ball{a30}
G(\b r, t| \b r', t')= e^{-\sigma t/2}\mathcal{G}(\b r, t| \b r', t')
\gal
is a Green's function of the Klein-Gordon equation with imaginary mass $m=i\sigma/2$
\ball{a32}
&-\nabla^2  \mathcal{G}+\partial_t^2 \mathcal{G}+ m^2 \mathcal{G}=e^{\sigma t'/2}\delta(t-t')\delta(\b r-\b r')\,.
\gal
The corresponding retarded Green's function in the coordinate representation reads (see e.g.\ \cite{MF})
\ball{a33}
\mathcal{G}(\b r, t| \b r', t')=&\frac{1}{4\pi}e^{\frac{1}{2}\sigma t' }\left\{ \frac{\delta(t-t'-R)}{R}\right.\nonumber\\
&\left.-\frac{m}{\sqrt{(t-t')^2-R^2}}J_1\left(m\sqrt{(t-t')^2-R^2}\right)\theta (t-t'-R)\right\}\theta(t-t')\,.
\gal
Eqs.~\eq{a30} and \eq{a33}  furnish the retarded Green's function for the original Eq.~\eq{a28}:
\begin{subequations}\label{a34}
\bal
G(\b r, t| \b r', t')&=G_a(\b r, t| \b r', t')+G_b(\b r, t| \b r', t')\\
G_a(\b r, t| \b r', t')&= \frac{1}{4\pi}e^{-\frac{1}{2}\sigma(t- t')}\frac{\delta(t-t'-R)}{R}\theta(t-t')\label{a34a}\\
G_b(\b r, t| \b r', t')&=\frac{1}{4\pi}e^{-\frac{1}{2}\sigma(t- t')}\frac{\sigma/2}{\sqrt{(t-t')^2-R^2}}I_1\left(\frac{\sigma}{2}\sqrt{(t-t')^2-R^2}\right)\theta (t-t'-R)\theta(t-t')\,.\label{a34b}
\gal
\end{subequations}
 We separated the Green's function into a sum of the two terms: the original pulse $G_a$ and the wake $G_b$ created by the currents induced in the plasma. The exponential factor $\exp[-\sigma(t-t')/2]$ indicates the decrease of the field strength due to the work done by the field on the electric currents in the plasma.

%%%%%%%%%%%%%%%%%%%%
\section{Solution for the static plasma}\label{sec:c}

The particular solution to \eq{a25}, namely the one induced by the external currents, is given by
\ball{a38}
 \b A^{(0)}(\b r, t)= \int G(\b r, t| \b r', t')\b j (\b r', t')d^3r' dt'\,,
 \gal
 where the retarded Green's function is given by \eq{a34}.
 Since the retarded Green's function breaks up into two physically meaningful terms we compute and analyze each term independently. 
 
\subsection{The pulse field}

The argument of the delta function in $G_a$ vanishes when $t-t'= |\b r-vt' \unit z|$. The corresponding retarded time $t'$ satisfying $t>t'$ reads 
\ball{a42} 
t'=t_0= \gamma^2\left( t-vz-\sqrt{(z-vt)^2+b^2/\gamma^2}\right)\,.
\gal
Writing 
\ball{a44}
\delta(t-t'-R)= \frac{\delta(t'-t_0) (t-t_0)}{\sqrt{(z-vt)^2+b^2/\gamma^2}}\,
\gal
and denoting $\xi = vt-z$ we find 
\ball{a46}
\b A_a^{(0)}(\b r, t)= \frac{ev\unit z}{4\pi}\frac{1}{\sqrt{\xi^2+b^2/\gamma^2}}\exp\left\{-\frac{\sigma\gamma^2}{2} \left(-v\xi+\sqrt{\xi^2+b^2/\gamma^2}\right)\right\}\,.
\gal
It is readily seen that as $\sigma\to 0$ this term reproduces the vector potential of a charge uniformly moving in vacuum. The  magnetic field corresponding to the vector potential \eq{a46} is given by
\bal
\b B_a^{(0)}&= -\frac{\partial  A_{az}^{(0)}}{\partial b}\unit \phi\label{a50}\\
& = \frac{ev}{4\pi}\unit \phi \left\{ \frac{\sigma b/2 }{\xi^2+b^2/\gamma^2}+ \frac{b}{\gamma^2 [\xi^2+b^2/\gamma^2]^{3/2}}
\right\}\exp\left\{-\frac{\sigma\gamma^2}{2} \left(-v\xi+\sqrt{\xi^2+b^2/\gamma^2}\right)\right\}\,.
\label{a51}
\gal
The first term in the curly brackets dominates when $\sqrt{\xi^2+b^2/\gamma^2}\gg 1/\sigma\gamma^2\sim 10^{-5}$~fm. Assuming that this is the case, \eq{a51} simplifies in the limit $b/\gamma \ll \xi$ yielding the ``diffusion approximation"
\ball{a53} 
\b B_a^{(0)}\approx \frac{ev}{8\pi}\unit \phi \frac{\sigma b }{\xi^2}e^{-\frac{\sigma\xi}{2(1+v)}}e^{-\frac{b^2\sigma}{4\xi}}\,,\quad \xi>0\,.
\gal
 Clearly, the second exponential factor in \eq{a53} can be dropped at later times $\xi\gg b^2\sigma/4\sim 0.5$~fm. 

The expression for the magnetic field was previously derived by one of us in \cite{Tuchin:2013apa} (see Eq.~(7) there) and, unlike \eq{a51},  is represented in a form of a one-dimensional integral. Both formulas reduce to  \eq{a53} in the diffusion approximation. 

%%%%%%%
\subsection{The wake field}

It has been tacitly assumed in \cite{Tuchin:2013apa} that the wake term is small. Using the Green's function \eq{a34b} we can compute this term explicitly:
\ball{a58}
\b A_b^{(0)}(\b r, t)=\frac{e\unit z}{4\pi}\frac{\sigma  v}{2}\int_{-\infty}^{t_0}
\frac{e^{-\sigma(t-t')/2}}{\sqrt{(t-t')^2-b^2-(z-vt')^2}}I_1\left( \frac{\sigma}{2}\sqrt{(t-t')^2-b^2-(z-vt')^2}\right)dt'\,.
\gal
It is useful to introduce a new integration variable $\lambda$ such that
\ball{a60}
t'= \gamma^2\left( t-vz-\sqrt{(z-vt)^2+(b^2+\lambda^2)/\gamma^2}\right)\,.
\gal
It is straightforward  to check that this implies 
\ball{a62}
\lambda^2= (t-t')^2-b^2-(z-vt')^2\,.
\gal
The vector potential  \eq{a58} can now be represented as 
\ball{a64}
\b A_b^{(0)}(\b r, t)=\frac{e\unit z}{4\pi}\frac{\sigma  v}{2}\int^{\infty}_{0}
\frac{d\lambda\, I_1\left( \frac{\sigma}{2}\lambda\right)}{\sqrt{\xi^2+(b^2+\lambda^2)/\gamma^2}}\exp\left\{ -\frac{\sigma\gamma^2}{2} \left(-v\xi+\sqrt{\xi^2+(b^2+\lambda^2)/\gamma^2}\right)\right\}\,.
\gal
The main contribution to this integral comes from the integration region $\sqrt{\gamma^2\xi^2+b^2}\ll \lambda \ll 2/\sigma\gamma$ where the integrand is approximately constant. At smaller $\lambda$'s it vanishes as $\sim\lambda$, while at larger $\lambda$'s it is exponentially suppressed. Thus, we can approximate the integral in \eq{a64} as 
\bal
\b A_b^{(0)}(\b r, t)&\approx \frac{e\unit z}{4\pi}\frac{\sigma  v}{2}\int^{\infty}_{0}
\frac{d\lambda\, \frac{1}{2} \frac{\sigma}{2}\lambda}{\sqrt{\xi^2+(b^2+\lambda^2)/\gamma^2}}\exp\left\{ -\frac{\sigma\gamma^2}{2} \left(-v\xi+\sqrt{\xi^2+(b^2+\lambda^2)/\gamma^2}\right)\right\}\nonumber\\
&=\frac{e\unit z}{4\pi}\frac{\sigma  v}{4}\exp\left\{-\frac{\sigma\gamma^2}{2} \left(-v\xi+\sqrt{\xi^2+b^2/\gamma^2}\right)\right\}\,.\label{a67}
\gal
Using \eq{a50} we derive the magnetic field 
\ball{a70}
\b B_b^{(0)}(\b r, t) =\frac{e\unit \phi}{4\pi}\frac{\sigma^2  v b}{4}\frac{1}{\sqrt{\xi^2+b^2/\gamma^2}}\exp\left\{-\frac{\sigma\gamma^2}{2} \left(-v\xi+\sqrt{\xi^2+b^2/\gamma^2}\right)\right\}\,.
\gal
Comparing \eq{a67} and \eq{a46} we conclude that the contribution of the wake to the retarded Greens function \eq{a34}  is small in the phenomenologically relevant region $\sqrt{\xi^2+b^2/\gamma^2}\ll 4/\sigma\sim 10^2$~fm. However, it dominates in the opposite limit, i.e.\ at very late times. 

%%%%%%%%%%%%%
\subsection{Diffusion approximation} \label{sec:d}

It is instructive to derive Eq.~\eq{a53} directly  from \eq{a28} as has been done in \cite{Tuchin:2015oka}. The diffusion approximation in \eq{a28} amounts to the assumption that  $\partial_z^2-\partial_t^2\sim k_z^2/\gamma^2\ll k_\bot^2, \sigma k_z$. In this case the retarded Green's function $G_\mathcal{D}(\b r, t| \b r', t')$ obeys the equation
\ball{d1}
&-\nabla_\bot^2  G_\mathcal{D}+ \sigma\partial_t G_\mathcal{D}=\delta(t-t')\delta(\b r-\b r')\,.
\gal
Its solution is 
\ball{d3}
G_\mathcal{D}(\b r, t| \b r', t')= \int \frac{d^3p}{(2\pi)^3}\int_{-\infty}^\infty \frac{d\omega}{2\pi }\frac{e^{-i\omega (t-t') +i\b p\cdot (\b r-\b r')}}{p_\bot^2-i\omega \sigma}= \frac{1}{4\pi t}\delta(z-z')\theta(t-t') e^{-\frac{\sigma (\b r_\bot-\b r_\bot')^2}{4(t-t')}}\,.
\gal
Employing \eq{a23} and \eq{a38} one derives
\ball{d5}
\b A^{(0)}(\b r, t) = \frac{e\unit z}{4\pi(t-z/v)}e^{-\frac{\sigma b^2}{4(t-z/v)}}\theta(t-z/v)\,,
\gal
which yields \eq{a53} for $\xi \ll 4/\sigma$\,.

%%%%%%%%%%%%%%%%%%%%%%%
\section{Solution for the expanding plasma}\label{sec:e}

\subsection{Contribution of the plasma flow}

\begin{figure}[ht]
      \includegraphics[height=8cm]{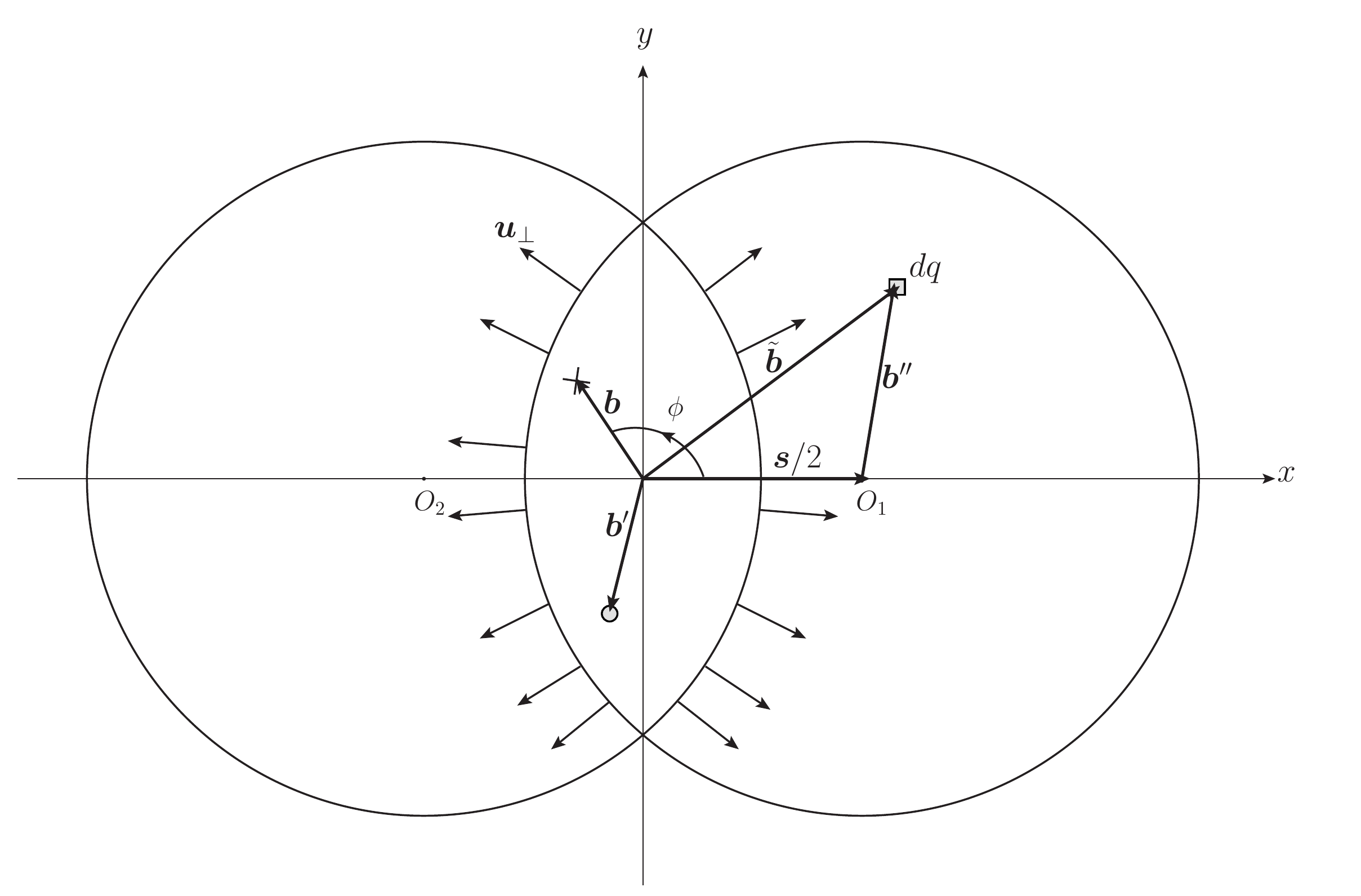} 
  \caption{The geometry of the heavy-ion collisions in the transverse plane. The two heavy-ion remnants (big circles) move in opposite directions along the $z$-axis, see \fig{geom-1}. The element of charge $dq$ is located at the same $z$ as an ion remnant (i.e.\ it is not inside the plasma). Its projection on the transverse plane is  depicted by the square. The small circle indicates the element of plasma moving with velocity $\b u$. The observation point is denoted by the $+$ symbol. The impact parameter $\b s$ points from  one nuclear center to another one. 
     }
\label{geom-2}
\end{figure}

Now we turn to Eq.~\eq{a26} that takes the plasma flow into account. Suppose that a point source is moving along the trajectory $z=vt$, $x=\tilde x$, $y=\tilde y$, where $\tilde x$ and $\tilde y$ are constants, see \fig{geom-2}. Denote by $\tilde{\b r}$ a vector with components $\tilde x,\tilde y,z$ and let $\tilde{\b b}$ be its transverse part.  The magnetic field created by this charge in the stationary plasma is then given by \eq{a51} and \eq{a70} with the replacement $ b \to |\b b- \tilde{\b b}|$; denote it as  $\b B^{(0)}(\b r-\tilde{\b r},t)$. The solution to \eq{a26} can be written right away using the Green's function as
\ball{b1}
\b A^{(1)}(\b r, t|\tilde{\b r})=\sigma \int G_a(\b r, t| \b r', t') \b u(\b r',t') \times \b B^{(0)}_a(\b r'-\tilde{\b r}, t')d^3r' dt'\,.
\gal
The contribution of the wake is neglected as per the results of the previous section. 

The longitudinal expansion of QGP is usually described by  the Bjorken model \cite{Bjorken:1982qr} in which the flow velocity in the lab frame is given by 
\ball{b15}
\b u(\b r, t) =  \frac{\b z}{t}\,.
\gal
Since the plasma velocity is non-vanishing only in the forward light-cone, i.e.\ $\b u^2\le 1$,  the integral in \eq{b1} is restricted to the region $|z'|\le t'$. Using $t'= t-R$ this implies that the integral over $z'$ runs between the following limits:
\ball{b17}
-\frac{t^2-z^2-(\b b-\b b')^2}{2(t+z)}\le z' \le \frac{t^2-z^2-(\b b-\b b')^2}{2(t-z)}\,.
\gal
In fact, the applicability of the Bjorken model is restricted to the central plateau region in the inclusive particle spectrum at a given energy. If $2Y$ is the extent of the plateau in rapidity, then $|\b u|\le \tanh Y$. For a conservative estimate of the flow correction we set $Y$ to infinity, which yields \eq{b17}.  

A more sophisticated blast wave model \cite{Siemens:1978pb,Teaney:2000cw,Kolb:2000fha} takes the transverse flow into account
\ball{b19}
\b u(\b r, t) = \frac{u_o}{R_o}\b b\,\theta(R_o-b)+  \frac{\b z}{t}\,,
\gal
where $u_o$ and $R_o$ are parameters fitted to the experimental data. We use $R_o=7.5$~fm, $u_o=0.55$ from \cite{Teaney:2003kp}. This time, restriction to the forward light-cone $\b u^2(\b r', t')\le 1$ reads
\ball{b21}
\left(\frac{u_o b'}{R_o}\right)^2+\left(\frac{z'}{t-R}\right)^2\le 1\,.
\gal

%%%%%%
%%%%%

%%%%%%%%%%%%%%%%%%%%%%%
\subsection{Initial conditions}\label{sec:f}

Thus far we assumed that a particle moves in plasma all the way from $t=-\infty$. In fact, a physical scenario more relevant for relativistic heavy-ion collisions is that the valence charges move in vacuum until a certain time $\tau$ when the plasma emerges. We neglect the finite 
thermalization time. Let the initial conditions be
\ball{f1}
\b A(\b r, \tau)= \bm{\mathcal{A}}(\b r)\,,\quad \frac{\partial \b A(\b r, \tau)}{\partial t}= \bm{\mathcal{V}}(\b r)\,,
\gal
where $\bm{\mathcal{A}}$ and $\bm{\mathcal{V}}$ are determined by the field that existed before the plasma emergence at $t=\tau$ \cite{Tuchin:2015oka}.
Then, the solution to \eq{a25} can be written as
\begin{subequations}
\bal
\b A^{(0)}(\b r, t)=& \int_\tau^{t+}dt' \int d^3r' \b j(\b r',t')G(\b r, t|\b r',t')\label{f4}\\
&+ \int d^3r' \left\{ \sigma \bm{\mathcal{A}}(\b r')+\bm{\mathcal{V}}(\b r')\right\} 
G(\b r, t|\b r',\tau)\label{f5}\\
&-\int d^3r' \bm{\mathcal{A}}(\b r')\frac{\partial}{\partial t'}G(\b r, t|\b r',\tau)\,.\label{f6}
\gal
\end{subequations}
The initial conditions \eq{f5} and \eq{f6} are satisfied at the leading order. Since they are independent of the plasma flow, we are not going to be concerned with them anymore in this paper. Thus, the solution to \eq{a26} takes form 
\ball{f8}
\b A^{(1)}(\b r, t|\tilde{\b r})=\sigma \int_\tau^{t+}dt' \int d^3r' G(\b r, t|\b r',t')\,\b u(\b r',t')\times \b B^{(0)}(\b r'- \tilde{\b r}, t')\,.
\gal
The initial time is chosen to be $\tau=0.2$~fm/$c$ in accordance with the phenomenological models of relativistic heavy-ion collisions \cite{Kolb:2000fha,Teaney:2000cw}. 

%%%%%%%%%%%%%
\subsection{Magnetic field of a nucleus}\label{sec:g}

The total field created by a nucleus is
\ball{g1}
\b A_\text{nucl}(\b r, t) =\int \rho (\b r'')\b A^{(0)}(\b b- \tilde{\b b},z-\tilde z,  t)d^3 r''+ \int \rho (\b r'')\b A^{(1)}(\b r, t|\tilde{\b b},\tilde z)d^3 r''\,,
\gal
where we slightly modified the notation by replacing $\tilde{\b r}$ with $\tilde{\b b},\tilde z$ in the vector potential argument. 
In the laboratory frame, the proton distribution in the nucleus in the $z$-direction is very narrow with average coordinate $\tilde z= vt$ depending on the direction of motion. Assuming that the nuclear density $\rho$  is constant throughout  the nucleus of radius $R_A$ and using \fig{geom-2} one can compute the vector potential as
\bal
\b A^{(0)}_\text{nucl}(\b r, t) &= 2\int \rho  \sqrt{R_A^2-(b'')^2}\b A^{(0)}(\b b - \b b''-\b s/2,z- vt, t)d^2  b''\label{g3}\\
\b A^{(1)}_\text{nucl}(\b r, t) &= 2\int \rho  \sqrt{R_A^2-(b'')^2}\b A^{(1)}(\b r, t|\ \b s/2+\b b'', vt)d^2  b''\,.\label{g4}
\gal
The nuclear density is normalized as $\rho (4\pi /3)R_A^3 = Z$, where $Ze$ is the nucleus electric charge. The contribution of another heavy-ion can be calculated by simply replacing $\b v\to -\b v$. In the figures below we show only the single nucleus contribution.

It follows from \eq{g3} that the magnetic field created by a single nucleus in a stationary plasma is 
\ball{g5}
\b B^{(0)}_\text{nucl}(\b r, t) = 2\int \rho  \sqrt{R_A^2-(b'')^2}\b B_a^{(0)}(\b b - \b b''-\b s/2,z- vt, t)d^2  b''\,,
\gal
where only the pulse contribution \eq{a51} is taken into account, whereas the wake contribution \eq{a70} is neglected. Since $\b A^{(0)}_\text{nucl}$ is directed along the $z$-axis, the corresponding magnetic field $\b B^{(0)}_\text{nucl}$ is circularly polarized in the $\unit \phi$ direction with respect to the nuclear center $O_1$ (or $O_2$). It is related to the radial $\unit b$ and the polar $\unit \varphi$ unit vectors of the cylindrical coordinate system defined with respect to the ``lab" reference frame shown in \fig{geom-2} as 
\ball{b7}
\unit \phi = \unit b\sin (\phi-\zeta)+ \unit \varphi \cos(\phi- \zeta)\,, 
\gal
where $\zeta$ given by  
\ball{g9}
\cot \zeta = \frac{b\cos\phi-s/2}{b\sin\phi}
\gal
is the angle between the vector pointing from $O_1$ to the observation point and the $x$-axis. 
The correction \eq{g4} due to the plasma expansion can be written down using \eq{g5} as
\ball{g11}
\b A^{(1)}_\text{nucl}(\b r, t) = \sigma \int _\tau ^{t+}dt'\int d^3r' G_a(\b r, t|\b r', t')\b u(\b r', t')\times \b B^{(0)}_\text{nucl}(\b r', t')\,.
\gal
In view of \eq{b19}, this equation indicates  that the longitudinal expansion of plasma induces the transverse  $\unit \varphi$ and $\b b$ components of the vector potential, while the transverse expansion induces a small $z$-correction to the vector potential. Moreover, according to \eq{b7}, $A^{(1)}_\varphi/A^{(1)}_b = -\tan(\phi-\zeta)$.

%%%%
\begin{figure}[ht]
\begin{tabular}{cc}
      \includegraphics[height=5cm]{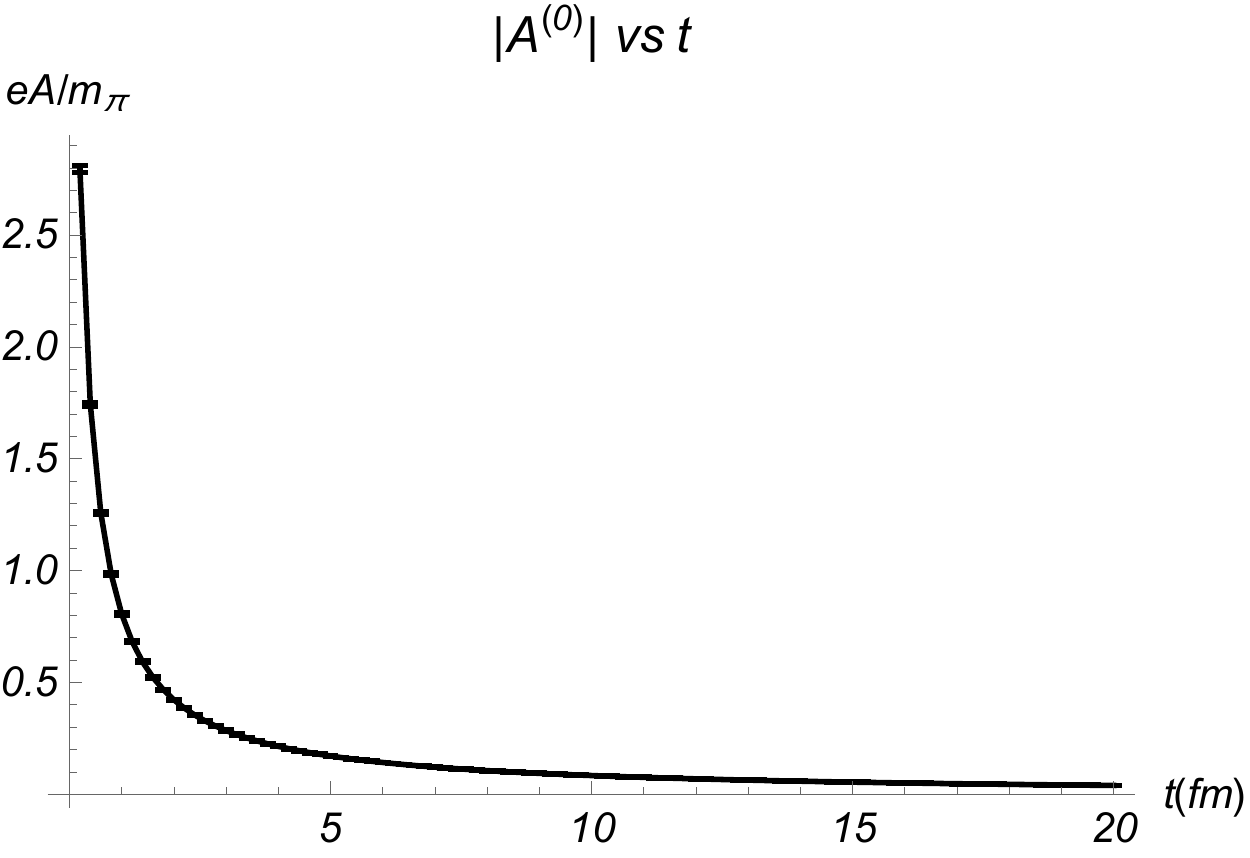} &
      \includegraphics[height=5cm]{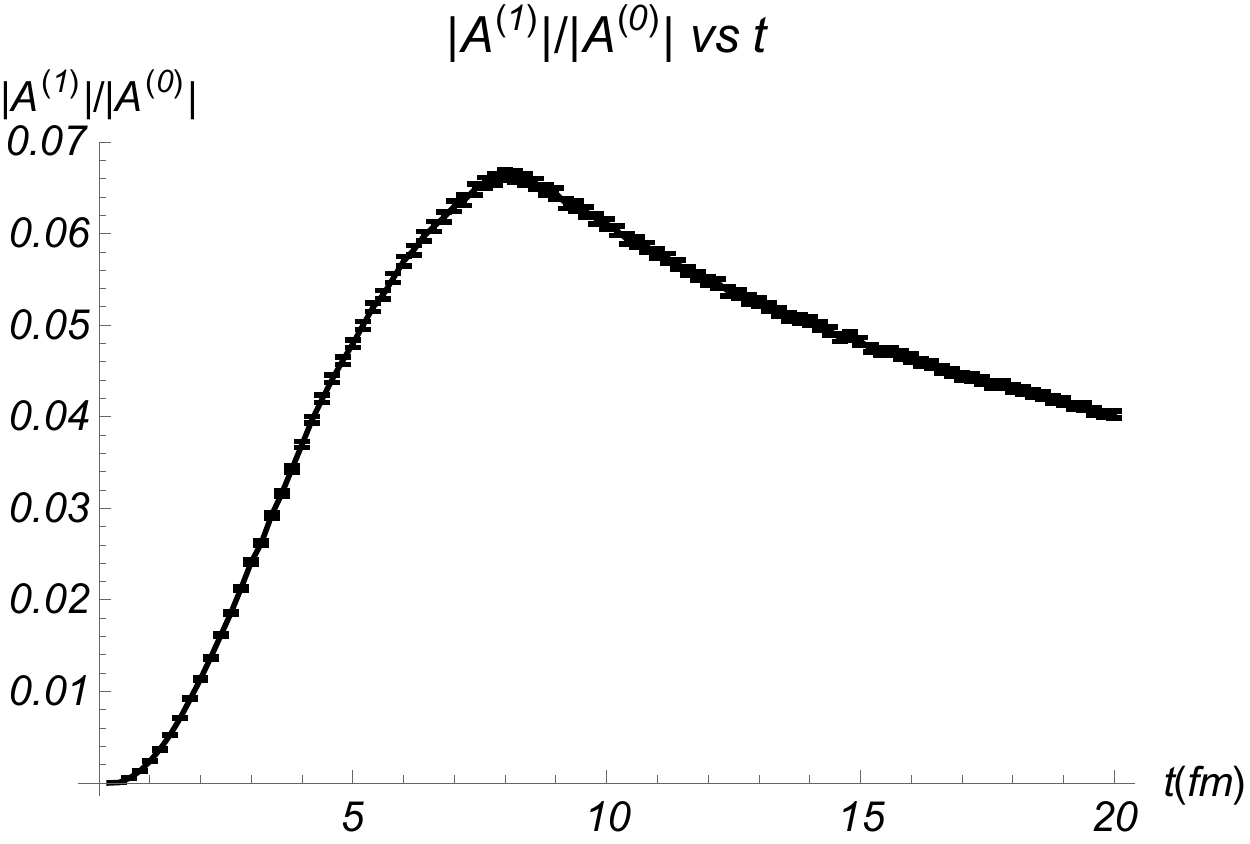}
      \end{tabular}
  \caption{The vector potential $\b A= \b A^{(0)}+ \b A^{(1)}$ created at a representative point $z=0$, $b=1$~fm, $\phi=\pi/6$ (see \fig{geom-2}) in QGP  by a remnant of the gold ion moving with the boost-factor $\gamma=100$ ($\sqrt{s}=0.2$ TeV) and impact parameter $|\b s|=3$~fm. Left panel: vector potential  $\b A^{(0)}$ in the non-expanding plasma. Right panel: the relative contribution of  the plasma expansion.   The plasma emerges at $\tau = 0.2$~fm/$c$.}
\label{Numerics}
\end{figure}
%%%%%

In the left panel of \fig{Numerics} we show the time-dependence of the vector potential in the stationary plasma $\b A^{(0)}$ at a representative point indicated in the caption. This  calculation agrees with the previous results \cite{Tuchin:2013apa}. It is seen that the magnetic field appears at $t=\tau=0.2$~fm/$c$ because we assumed that QGP emerges at that time. It is important to mention that in this calculation we do not consider the contributions from the fields that existed at $t<\tau$. They are given by Eqs.~\eq{f5} and \eq{f6} and are not affected by the plasma flow, even though they give a significant contribution to  $\b A^{(0)}$ as shown in \cite{Tuchin:2015oka}. 

In the right panel of \fig{Numerics} we show the  time-dependence of the ratio  $ A^{(1)}/A^{(0)}$ at a representative point inside QGP, which illustrates the relative significance of the plasma expansion in the magnetic field calculations. The main observation is that the relative contribution of the plasma expansion is below 10\%. With this accuracy, the plasma expansion effect on the magnetic field can be safely neglected. 

%%%%%%
\begin{figure}[ht]
      \includegraphics[height=5cm]{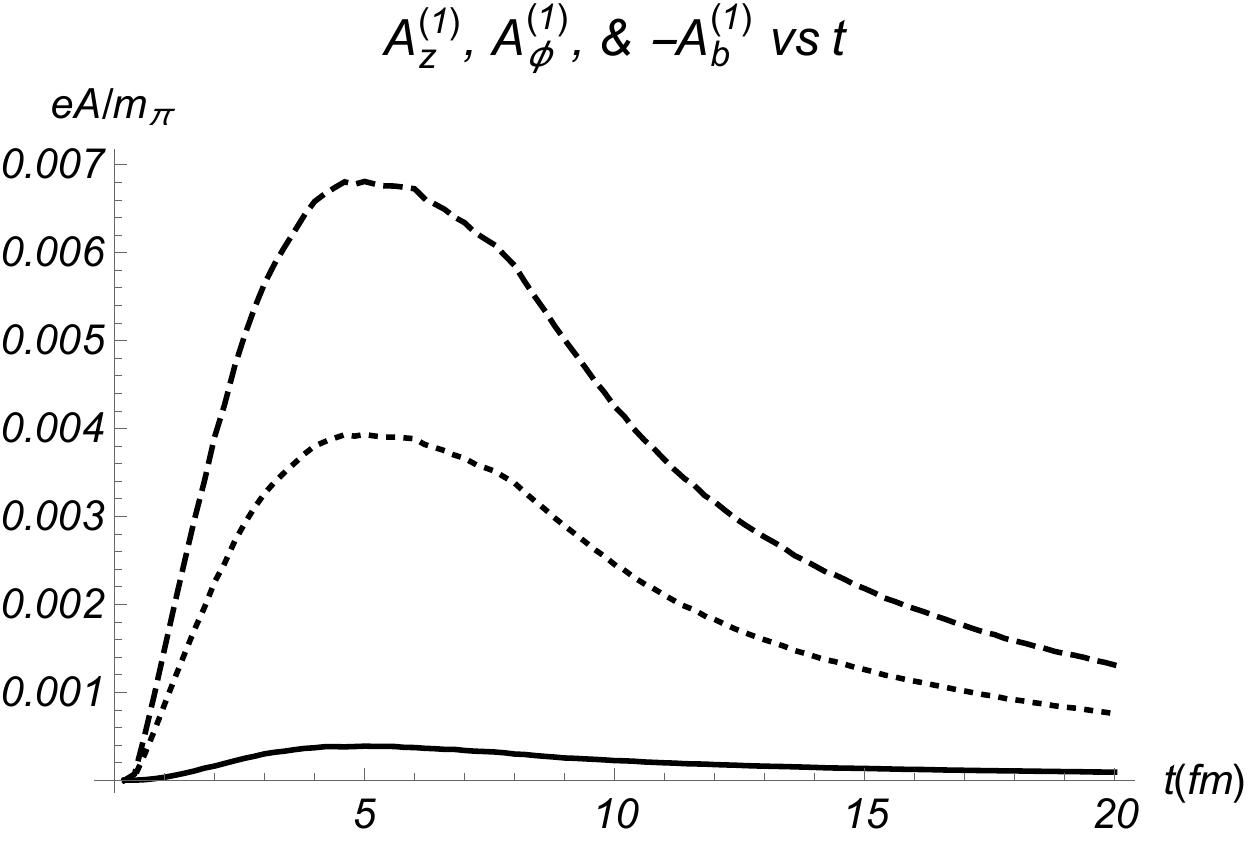} 
  \caption{Dotted line: $A_z^{(1)}$, dashed line: $A_\phi^{(1)}$, solid line $-A_b^{(1)}$ components of the correction $\b A^{(1)}$ to the vector potential (in units of $m_\pi/e$) due to the plasma expansion. The geometric and kinematic parameters are the same as in \fig{Numerics}. The cylindrical coordinates are defined with respect to the $z$-axis of \fig{geom-2}, which is the  lab frame for heavy-ion collisions.   }
\label{Numerics-2}
\end{figure}
%%%%%

\fig{Numerics-2} shows the components of the correction to the vector potential due to the plasma expansion. The vector potential in the stationary plasma always points in the direction of the external charge motion ($\pm \unit z$-directions) generating the total magnetic field as a superposition of the circularly polarized fields of the individual charges. In contrast,  flow of plasma generates additional components of the vector potential in the transverse plane.

The vector potentials shown in \fig{Numerics} and \fig{Numerics-2} is produced by a relativistic heavy-ion in a single event. We assumed that the electric charge distribution in the rest frame is uniform across the nucleus. Using a more accurate Woods-Saxon distribution gives a tiny correction. Many transport models treat heavy ion as a collection of electric charges of finite radius that are randomly distributed according to a given average charge distribution. This produces large event-by-event fluctuations of charge positions, which in turn induces large event-by-event fluctuations of electromagnetic field \cite{Bzdak:2011yy}. However, it was shown in  \cite{Zakharov:2017gkb} that the quantum treatment of the nuclear electric charge distribution yields  fluctuations which are roughly an order of magnitude smaller than the flow contribution. In view of this observation we neglected the event-by-event fluctuations in this paper.

%%%%%%%%%%%%%
\section{Summary}\label{sec:s}

We computed the effect of the QGP expansion on the magnetic field created inside the plasma by external valence charges of the heavy-ion remnants. Our main assumption is that the plasma flow is not affected by the magnetic field and is given by the phenomenological blast-wave model. We treated the effect of plasma flow as a perturbation of the magnetic field in a stationary plasma. The result shown in \fig{Numerics} indicates that the contribution of the plasma flow to the magnetic field is less than 10\%. Our main conclusion is that  there is no urgent need to solve the comprehensive MHD equations in order to describe the QGP dynamics at present energies, unless one wishes to reach precision of about 10\%. It is a very good approximation, on the one hand, to study QGP in the  background electromagnetic field generated by external sources and, on the other hand,  to investigate the dynamics of magnetic field in the background plasma. 

Since in this paper we focused on the contribution of plasma flow to the magnetic field of external charges, we disregarded  the magnetic field created by the fields that existed before the plasma emergence. However, in phenomenological applications they certainly have to be taken into account as argued in \cite{Tuchin:2015oka}. Incidentally, we observed that the diffusion approximation used in \cite{Tuchin:2015oka} to analyze the initial conditions is quite reasonable. 

In our previous calculations of magnetic field we always tacitly neglected the wake produced by the currents induced in plasma. In \sec{sec:c} we derived the analytic expressions for the pulse and wake fields, given by \eq{a51} and \eq{a70} respectively, and argued that the wake field is indeed negligible in the phenomenologically relevant regime due to the smallness of the electrical conductivity as compared to the inverse QGP lifetime.

Our paper paves the road to a comprehensive computation of electromagnetic field with quantum sources, whose importance was demonstrated in  \cite{Holliday:2016lbx,Peroutka:2017esw}. The fact that the flow of plasma and the wake effects are but small corrections is enormous simplification of the MHD equations. Computing such a field with the appropriate initial conditions is  the subject of our forthcoming paper.

%%%%%%%%%%%%%%%%%%%%%%%%%%%%%%%%
\acknowledgments
  This work  was supported in part by the U.S. Department of Energy under Grant No.\ DE-FG02-87ER40371.

%% appendix 
%\appendix
%\section{}\label{appA}

%%%%%%%%%%%%%%%%%%%%%%%%%%%%%%%%%%%%%


\begin{thebibliography}{80}
  
  

  

  
  



%%%%%%  begin MHD  %%%%%%%%%%%%%%%%

  %\cite{Mohapatra:2011ku}
\bibitem{Mohapatra:2011ku} 
  R.~K.~Mohapatra, P.~S.~Saumia and A.~M.~Srivastava,
  ``Enhancement of flow anisotropies due to magnetic field in relativistic heavy-ion collisions,''
  Mod.\ Phys.\ Lett.\ A {\bf 26}, 2477 (2011)
%  doi:10.1142/S0217732311036711
  %[arXiv:1102.3819 [hep-ph]].
  %%CITATION = doi:10.1142/S0217732311036711;%%
  %19 citations counted in INSPIRE as of 08 Sep 2017
  

%\cite{Tuchin:2011jw}
\bibitem{Tuchin:2011jw} 
  K.~Tuchin,
  ``On viscous flow and azimuthal anisotropy of quark-gluon plasma in strong magnetic field,''
  J.\ Phys.\ G {\bf 39}, 025010 (2012)
  %doi:10.1088/0954-3899/39/2/025010
  %[arXiv:1108.4394 [nucl-th]].
  %%CITATION = doi:10.1088/0954-3899/39/2/025010;%%
  %30 citations counted in INSPIRE as of 08 Sep 2017

  

 %\cite{Roy:2015kma}
\bibitem{Roy:2015kma} 
  V.~Roy, S.~Pu, L.~Rezzolla and D.~Rischke,
  ``Analytic Bjorken flow in one-dimensional relativistic magnetohydrodynamics,''
  Phys.\ Lett.\ B {\bf 750}, 45 (2015)
  %doi:10.1016/j.physletb.2015.08.046
  %[arXiv:1506.06620 [nucl-th]].
  %%CITATION = doi:10.1016/j.physletb.2015.08.046;%%
  %11 citations counted in INSPIRE as of 22 Nov 2016

  
  %\cite{Roy:2015coa}
\bibitem{Roy:2015coa} 
  V.~Roy and S.~Pu,
  ``Event-by-event distribution of magnetic field energy over initial fluid energy density in $\sqrt{s_{\rm NN}}$= 200 GeV Au-Au collisions,''
  Phys.\ Rev.\ C {\bf 92}, 064902 (2015)
  %doi:10.1103/PhysRevC.92.064902
  %[arXiv:1508.03761 [nucl-th]].
  %%CITATION = doi:10.1103/PhysRevC.92.064902;%%
  %18 citations counted in INSPIRE as of 08 Sep 2017

%\cite{Pu:2016ayh}
\bibitem{Pu:2016ayh} 
  S.~Pu, V.~Roy, L.~Rezzolla and D.~H.~Rischke,
  ``Bjorken flow in one-dimensional relativistic magnetohydrodynamics with magnetization,''
  Phys.\ Rev.\ D {\bf 93}, no. 7, 074022 (2016)
  %doi:10.1103/PhysRevD.93.074022
  %[arXiv:1602.04953 [nucl-th]].
  %%CITATION = doi:10.1103/PhysRevD.93.074022;%%
  %7 citations counted in INSPIRE as of 22 Nov 2016  

%\cite{Inghirami:2016iru}
\bibitem{Inghirami:2016iru} 
  G.~Inghirami, L.~Del Zanna, A.~Beraudo, M.~H.~Moghaddam, F.~Becattini and M.~Bleicher,
  ``Numerical magneto-hydrodynamics for relativistic nuclear collisions,''
  Eur.\ Phys.\ J.\ C {\bf 76}, no. 12, 659 (2016)
  %doi:10.1140/epjc/s10052-016-4516-8
  %[arXiv:1609.03042 [hep-ph]].
  %%CITATION = doi:10.1140/epjc/s10052-016-4516-8;%%
  %2 citations counted in INSPIRE as of 07 Mar 2017
  
 %\cite{Pu:2016bxy}
\bibitem{Pu:2016bxy} 
  S.~Pu and D.~L.~Yang,
  ``Transverse flow induced by inhomogeneous magnetic fields in the Bjorken expansion,''
  Phys.\ Rev.\ D {\bf 93}, no. 5, 054042 (2016)
  %doi:10.1103/PhysRevD.93.054042
  %[arXiv:1602.04954 [nucl-th]].
  %%CITATION = doi:10.1103/PhysRevD.93.054042;%%
  %14 citations counted in INSPIRE as of 08 Sep 2017
 
%\cite{Roy:2017yvg}
\bibitem{Roy:2017yvg} 
  V.~Roy, S.~Pu, L.~Rezzolla and D.~H.~Rischke,
  ``Effect of intense magnetic fields on reduced-MHD evolution in $\sqrt{s_{\rm NN}}$ = 200 GeV Au+Au collisions,''
  arXiv:1706.05326 [nucl-th].
  %%CITATION = ARXIV:1706.05326;%%
  %2 citations counted in INSPIRE as of 08 Sep 2017


%\cite{Das:2017qfi}
\bibitem{Das:2017qfi} 
  A.~Das, S.~S.~Dave, P.~S.~Saumia and A.~M.~Srivastava,
  ``Effects of magnetic field on the plasma evolution in relativistic heavy-ion collisions,''
  Phys.\ Rev.\ C {\bf 96}, no. 3, 034902 (2017)
  %doi:10.1103/PhysRevC.96.034902
  %[arXiv:1703.08162 [hep-ph]].
  %%CITATION = doi:10.1103/PhysRevC.96.034902;%%
  %6 citations counted in INSPIRE as of 08 Sep 2017
  
 %\cite{Greif:2017irh}
\bibitem{Greif:2017irh} 
  M.~Greif, C.~Greiner and Z.~Xu,
  ``Magnetic field influence on the early time dynamics of heavy-ion collisions,''
  Phys.\ Rev.\ C {\bf 96}, no. 1, 014903 (2017)
  %doi:10.1103/PhysRevC.96.014903
  %[arXiv:1704.06505 [hep-ph]].
  %%CITATION = doi:10.1103/PhysRevC.96.014903;%%
  %3 citations counted in INSPIRE as of 08 Sep 2017


%%%%%%  end MHD  %%%%%%%%%%%%%%%%



 
  
 %%%%  Electrical conductivity of QGP %%%%

 
  %\cite{Aarts:2007wj}
\bibitem{Aarts:2007wj}
  G.~Aarts, C.~Allton, J.~Foley, S.~Hands and S.~Kim,
  ``Spectral functions at small energies and the electrical conductivity in
  hot, quenched lattice QCD,''
  Phys.\ Rev.\ Lett.\  {\bf 99}, 022002 (2007)
 % [arXiv:hep-lat/0703008].
  %%CITATION = PRLTA,99,022002;%%

%\cite{Ding:2010ga}
\bibitem{Ding:2010ga} 
  H.-T.~Ding, A.~Francis, O.~Kaczmarek, F.~Karsch, E.~Laermann and W.~Soeldner,
  ``Thermal dilepton rate and electrical conductivity: An analysis of vector current correlation functions in quenched lattice QCD,''
  Phys.\ Rev.\ D {\bf 83}, 034504 (2011)
  %doi:10.1103/PhysRevD.83.034504
  %[arXiv:1012.4963 [hep-lat]].
  %%CITATION = doi:10.1103/PhysRevD.83.034504;%%
  %172 citations counted in INSPIRE as of 08 Sep 2017
   


  
  %\cite{Amato:2013oja}
\bibitem{Amato:2013oja} 
  A.~Amato, G.~Aarts, C.~Allton, P.~Giudice, S.~Hands and J.~I.~Skullerud,
  ``Transport coefficients of the QGP,''
  PoS LATTICE {\bf 2013}, 176 (2014)
  %[arXiv:1310.7466 [hep-lat]].
  %%CITATION = ARXIV:1310.7466;%%
  %11 citations counted in INSPIRE as of 08 Sep 2017
  
    
  %\cite{Cassing:2013iz}
\bibitem{Cassing:2013iz} 
  W.~Cassing, O.~Linnyk, T.~Steinert and V.~Ozvenchuk,
  ``On the electric conductivity of hot QCD matter,''
  Phys.\ Rev.\ Lett.\  {\bf 110}, 182301 (2013)
  %[arXiv:1302.0906 [hep-ph]].
  %%CITATION = ARXIV:1302.0906;%%
  %9 citations counted in INSPIRE as of 12 Dec 2013

  
  %\cite{Yin:2013kya}
\bibitem{Yin:2013kya} 
  Y.~Yin,
  ``Electrical conductivity of the quark-gluon plasma and soft photon spectrum in heavy-ion collisions,''
  Phys.\ Rev.\ C {\bf 90}, no. 4, 044903 (2014)
  %doi:10.1103/PhysRevC.90.044903
  %[arXiv:1312.4434 [nucl-th]].
  %%CITATION = doi:10.1103/PhysRevC.90.044903;%%
  %16 citations counted in INSPIRE as of 08 Sep 2017

%%%%  End of Electrical conductivity of QGP %%%% 


%%%%%%%% Begin TRANSPORT COEFF in B %%%%%%%%%%%%%

  %\cite{Chernodub:2009rt}
\bibitem{Chernodub:2009rt} 
  M.~N.~Chernodub, H.~Verschelde and V.~I.~Zakharov,
  ``Magnetic component of gluon plasma and its viscosity,''
  Nucl.\ Phys.\ Proc.\ Suppl.\  {\bf 207-208}, 325 (2010)
  %doi:10.1016/j.nuclphysbps.2010.10.079
  %[arXiv:0905.2520 [hep-ph]].
  %%CITATION = doi:10.1016/j.nuclphysbps.2010.10.079;%%
  %10 citations counted in INSPIRE as of 08 Sep 2017

  %\cite{Agasian:2011st}
\bibitem{Agasian:2011st} 
  N.~O.~Agasian,
  ``Low-energy theorems of QCD and bulk viscosity at finite temperature and baryon density in a magnetic field,''
  JETP Lett.\  {\bf 95}, 171 (2012)
  %doi:10.1134/S0021364012040029
  %[arXiv:1109.5849 [hep-ph]].
  %%CITATION = doi:10.1134/S0021364012040029;%%
  %7 citations counted in INSPIRE as of 08 Sep 2017
  
    %\cite{Nam:2013fpa}
\bibitem{Nam:2013fpa} 
  S.~i.~Nam and C.~W.~Kao,
  ``Shear viscosity of quark matter at finite temperature under an external magnetic field,''
  Phys.\ Rev.\ D {\bf 87}, no. 11, 114003 (2013)
  %doi:10.1103/PhysRevD.87.114003
  %[arXiv:1304.0287 [hep-ph]].
  %%CITATION = doi:10.1103/PhysRevD.87.114003;%%
  %12 citations counted in INSPIRE as of 08 Sep 2017

%\cite{Fukushima:2015wck}
\bibitem{Fukushima:2015wck} 
  K.~Fukushima, K.~Hattori, H.~U.~Yee and Y.~Yin,
  ``Heavy Quark Diffusion in Strong Magnetic Fields at Weak Coupling and Implications for Elliptic Flow,''
  Phys.\ Rev.\ D {\bf 93}, no. 7, 074028 (2016)
  %doi:10.1103/PhysRevD.93.074028
  %[arXiv:1512.03689 [hep-ph]].
  %%CITATION = doi:10.1103/PhysRevD.93.074028;%%
  %35 citations counted in INSPIRE as of 08 Sep 2017


  
  %\cite{Hattori:2016idp}
\bibitem{Hattori:2016idp} 
  K.~Hattori, K.~Fukushima, H.~U.~Yee and Y.~Yin,
  ``Heavy-Quark Diffusion Dynamics in Quark-Gluon Plasma under Strong Magnetic Fields,''
  arXiv:1611.00500 [hep-ph].
  %%CITATION = ARXIV:1611.00500;%%
  
  %\cite{Hattori:2016lqx}
\bibitem{Hattori:2016lqx} 
  K.~Hattori, S.~Li, D.~Satow and H.~U.~Yee,
  ``Longitudinal Conductivity in Strong Magnetic Field in Perturbative QCD: Complete Leading Order,''
  Phys.\ Rev.\ D {\bf 95}, no. 7, 076008 (2017)
 % doi:10.1103/PhysRevD.95.076008
  %[arXiv:1610.06839 [hep-ph]].
  %%CITATION = doi:10.1103/PhysRevD.95.076008;%%
  %11 citations counted in INSPIRE as of 08 Sep 2017

%\cite{Li:2016bbh}
\bibitem{Li:2016bbh} 
  S.~Li, K.~A.~Mamo and H.~U.~Yee,
  ``Jet quenching parameter of the quark-gluon plasma in a strong magnetic field: Perturbative QCD and AdS/CFT correspondence,''
  Phys.\ Rev.\ D {\bf 94}, no. 8, 085016 (2016)
  %doi:10.1103/PhysRevD.94.085016
  %[arXiv:1605.00188 [hep-ph]].
  %%CITATION = doi:10.1103/PhysRevD.94.085016;%%
  %14 citations counted in INSPIRE as of 08 Sep 2017
  
  %\cite{Hattori:2016cnt}
\bibitem{Hattori:2016cnt} 
  K.~Hattori and D.~Satow,
  ``Electrical Conductivity of Quark-Gluon Plasma in Strong Magnetic Fields,''
  Phys.\ Rev.\ D {\bf 94}, no. 11, 114032 (2016)
  %doi:10.1103/PhysRevD.94.114032
  %[arXiv:1610.06818 [hep-ph]].
  %%CITATION = doi:10.1103/PhysRevD.94.114032;%%
  %13 citations counted in INSPIRE as of 08 Sep 2017


%\cite{Li:2017tgi}
\bibitem{Li:2017tgi} 
  S.~Li and H.~U.~Yee,
  ``Shear Viscosity of Quark-Gluon Plasma in Weak Magnetic Field in Perturbative QCD: Leading Log,''
  arXiv:1707.00795 [hep-ph].
  %%CITATION = ARXIV:1707.00795;%%
  %2 citations counted in INSPIRE as of 08 Sep 2017
  

  
  %\cite{Hattori:2017qih}
\bibitem{Hattori:2017qih} 
  K.~Hattori, X.~G.~Huang, D.~H.~Rischke and D.~Satow,
  ``Bulk Viscosity of Quark-Gluon Plasma in Strong Magnetic Fields,''
  arXiv:1708.00515 [hep-ph].
  %%CITATION = ARXIV:1708.00515;%%


%\cite{Li:2017jwv}
\bibitem{Li:2017jwv} 
  Y.~Li and K.~Tuchin,
  ``Electrodynamics of dual superconducting chiral medium,''
  arXiv:1708.08536 [hep-ph].
  %%CITATION = ARXIV:1708.08536;%%

%%%%%%%% End TRANSPORT COEFF in B %%%%%%%%%%%%%



%%%%  Begin CALCS OF B IN HIC W/O CONDUCTIVITY %%%%%%

%\cite{Kharzeev:2007jp}
\bibitem{Kharzeev:2007jp}
  D.~E.~Kharzeev, L.~D.~McLerran and H.~J.~Warringa,
  ``The effects of topological charge change in heavy ion collisions: 'Event by
  event P and CP violation',''
  Nucl.\ Phys.\  A {\bf 803}, 227 (2008).
 % [arXiv:0711.0950 [hep-ph]].
  %%CITATION = NUPHA,A803,227;%%

%\cite{Skokov:2009qp}
\bibitem{Skokov:2009qp} 
  V.~Skokov, A.~Y.~Illarionov and V.~Toneev,
  ``Estimate of the magnetic field strength in heavy-ion collisions,''
  Int.\ J.\ Mod.\ Phys.\ A {\bf 24}, 5925 (2009)
  %doi:10.1142/S0217751X09047570
  %[arXiv:0907.1396 [nucl-th]].
  %%CITATION = doi:10.1142/S0217751X09047570;%%
  %413 citations counted in INSPIRE as of 22 Nov 2016

%\cite{Voronyuk:2011jd}
\bibitem{Voronyuk:2011jd} 
  V.~Voronyuk, V.~D.~Toneev, W.~Cassing, E.~L.~Bratkovskaya, V.~P.~Konchakovski and S.~A.~Voloshin,
  ``(Electro-)Magnetic field evolution in relativistic heavy-ion collisions,''
  Phys.\ Rev.\ C {\bf 83}, 054911 (2011)
  %doi:10.1103/PhysRevC.83.054911
  %[arXiv:1103.4239 [nucl-th]].
  %%CITATION = doi:10.1103/PhysRevC.83.054911;%%
  %150 citations counted in INSPIRE as of 22 Nov 2016
  
    %\cite{Ou:2011fm}
\bibitem{Ou:2011fm} 
  L.~Ou and B.~A.~Li,
  ``Magnetic effects in heavy-ion collisions at intermediate energies,''
  Phys.\ Rev.\ C {\bf 84}, 064605 (2011)
  %doi:10.1103/PhysRevC.84.064605
  %[arXiv:1107.3192 [nucl-th]].
  %%CITATION = doi:10.1103/PhysRevC.84.064605;%%
  %22 citations counted in INSPIRE as of 22 Nov 2016
  
     %\cite{Bzdak:2011yy}
\bibitem{Bzdak:2011yy} 
  A.~Bzdak and V.~Skokov,
  ``Event-by-event fluctuations of magnetic and electric fields in heavy ion collisions,''
  Phys.\ Lett.\ B {\bf 710}, 171 (2012)
  %doi:10.1016/j.physletb.2012.02.065
  %[arXiv:1111.1949 [hep-ph]].
  %%CITATION = doi:10.1016/j.physletb.2012.02.065;%%
  %156 citations counted in INSPIRE as of 22 Nov 2016
  
  %\cite{Bloczynski:2012en}
\bibitem{Bloczynski:2012en} 
  J.~Bloczynski, X.~G.~Huang, X.~Zhang and J.~Liao,
  ``Azimuthally fluctuating magnetic field and its impacts on observables in heavy-ion collisions,''
  Phys.\ Lett.\ B {\bf 718}, 1529 (2013)
  %doi:10.1016/j.physletb.2012.12.030
  %[arXiv:1209.6594 [nucl-th]].
  %%CITATION = doi:10.1016/j.physletb.2012.12.030;%%
  %43 citations counted in INSPIRE as of 22 Nov 2016
 
  
  %\cite{Deng:2012pc}
\bibitem{Deng:2012pc} 
  W.~T.~Deng and X.~G.~Huang,
  ``Event-by-event generation of electromagnetic fields in heavy-ion collisions,''
  Phys.\ Rev.\ C {\bf 85}, 044907 (2012)
  %doi:10.1103/PhysRevC.85.044907
  %[arXiv:1201.5108 [nucl-th]].
  %%CITATION = doi:10.1103/PhysRevC.85.044907;%%
  %186 citations counted in INSPIRE as of 22 Nov 2016
  
  %%%%  end CALCS OF B IN HIC W/O CONDUCTIVITY %%%%%%
  
%%%%%%%%   begin  CALCS OF B IN HIC WITH CONDUCTIVITY %%%%
 
   %\cite{Tuchin:2010vs}
\bibitem{Tuchin:2010vs}
  K.~Tuchin,
 ``Synchrotron radiation by fast fermions in heavy-ion collisions,''
  Phys.\ Rev.\  C {\bf 82}, 034904 (2010)
  [Erratum-ibid.\  C {\bf 83}, 039903 (2011)].
%  [arXiv:1006.3051 [nucl-th]].
  %%CITATION = PHRVA,C82,034904;%% 
 
  %\cite{Tuchin:2013ie}
\bibitem{Tuchin:2013ie} 
  K.~Tuchin,
  ``Particle production in strong electromagnetic fields in relativistic heavy-ion collisions,''
  Adv.\ High Energy Phys.\  {\bf 2013}, 490495 (2013)
  %[arXiv:1301.0099].
  %%CITATION = ARXIV:1301.0099;%%
  %74 citations counted in INSPIRE as of 26 Aug 2015
 
     %\cite{Tuchin:2013apa}
\bibitem{Tuchin:2013apa} 
  K.~Tuchin,
  ``Time and space dependence of the electromagnetic field in relativistic heavy-ion collisions,''
  Phys.\ Rev.\ C {\bf 88}, no. 2, 024911 (2013)
  %doi:10.1103/PhysRevC.88.024911
 % [arXiv:1305.5806 [hep-ph]].
  %%CITATION = doi:10.1103/PhysRevC.88.024911;%%
  %51 citations counted in INSPIRE as of 22 Nov 2016
 
  %\cite{Zakharov:2014dia}
\bibitem{Zakharov:2014dia} 
  B.~G.~Zakharov,
  ``Electromagnetic response of quark-gluon plasma in heavy-ion collisions,''
  Phys.\ Lett.\ B {\bf 737}, 262 (2014)
  %[arXiv:1404.5047 [hep-ph]].
  %%CITATION = ARXIV:1404.5047;%%
  %2 citations counted in INSPIRE as of 30 Oct 2014


  %\cite{Tuchin:2015oka}
\bibitem{Tuchin:2015oka} 
  K.~Tuchin,
  ``Initial value problem for magnetic fields in heavy ion collisions,''
  Phys.\ Rev.\ C {\bf 93}, no. 1, 014905 (2016)
  %doi:10.1103/PhysRevC.93.014905
  %[arXiv:1508.06925 [hep-ph]].
  %%CITATION = doi:10.1103/PhysRevC.93.014905;%%
  %8 citations counted in INSPIRE as of 22 Nov 2016
  
  %%%%%%%%   end  CALCS OF B IN HIC WITH CONDUCTIVITY %%%%
  
    
  %\cite{Bjorken:1982qr}
\bibitem{Bjorken:1982qr} 
  J.~D.~Bjorken,
  ``Highly Relativistic Nucleus-Nucleus Collisions: The Central Rapidity Region,''
  Phys.\ Rev.\ D {\bf 27}, 140 (1983).
  %doi:10.1103/PhysRevD.27.140
  %%CITATION = doi:10.1103/PhysRevD.27.140;%%
  %2753 citations counted in INSPIRE as of 08 Sep 2017
  
  %\cite{Kharzeev:2013ffa}
\bibitem{Kharzeev:2013ffa} 
  D.~E.~Kharzeev,
  ``The Chiral Magnetic Effect and Anomaly-Induced Transport,''
  Prog.\ Part.\ Nucl.\ Phys.\  {\bf 75}, 133 (2014)
  %doi:10.1016/j.ppnp.2014.01.002
  %[arXiv:1312.3348 [hep-ph]].
  %%CITATION = doi:10.1016/j.ppnp.2014.01.002;%%
  %59 citations counted in INSPIRE as of 25 Dec 2015
  
  %\cite{Huang:2015oca}
\bibitem{Huang:2015oca} 
  X.~G.~Huang,
  ``Electromagnetic fields and anomalous transports in heavy-ion collisions --- A pedagogical review,''
  Rept.\ Prog.\ Phys.\  {\bf 79}, no. 7, 076302 (2016)
  %doi:10.1088/0034-4885/79/7/076302
  %[arXiv:1509.04073 [nucl-th]].
  %%CITATION = doi:10.1088/0034-4885/79/7/076302;%%
  %60 citations counted in INSPIRE as of 08 Sep 2017  
  
  %\cite{Kharzeev:2015znc}
\bibitem{Kharzeev:2015znc} 
  D.~E.~Kharzeev, J.~Liao, S.~A.~Voloshin and G.~Wang,
  ``Chiral magnetic and vortical effects in high-energy nuclear collisions?A status report,''
  Prog.\ Part.\ Nucl.\ Phys.\  {\bf 88}, 1 (2016)
  %doi:10.1016/j.ppnp.2016.01.001
  %[arXiv:1511.04050 [hep-ph]].
  %%CITATION = doi:10.1016/j.ppnp.2016.01.001;%%
  %134 citations counted in INSPIRE as of 08 Sep 2017
   
%%%%%  begin EM fields and CME %%%%%

%\cite{Tuchin:2014iua}
\bibitem{Tuchin:2014iua} 
  K.~Tuchin,
  ``Electromagnetic field and the chiral magnetic effect in the quark-gluon plasma,''
  Phys.\ Rev.\ C {\bf 91}, no. 6, 064902 (2015)
  %doi:10.1103/PhysRevC.91.064902
  %[arXiv:1411.1363 [hep-ph]].
  %%CITATION = doi:10.1103/PhysRevC.91.064902;%%
  %20 citations counted in INSPIRE as of 22 Nov 2016
  
  %\cite{Manuel:2015zpa}
\bibitem{Manuel:2015zpa} 
  C.~Manuel and J.~M.~Torres-Rincon,
  ``Dynamical evolution of the chiral magnetic effect: Applications to the quark-gluon plasma,''
  Phys.\ Rev.\ D {\bf 92}, no. 7, 074018 (2015)
  %doi:10.1103/PhysRevD.92.074018
  %[arXiv:1501.07608 [hep-ph]].
  %%CITATION = doi:10.1103/PhysRevD.92.074018;%%
  %24 citations counted in INSPIRE as of 22 Nov 2016
  
  %\cite{Hirono:2015rla}
\bibitem{Hirono:2015rla} 
  Y.~Hirono, D.~Kharzeev and Y.~Yin,
  ``Self-similar inverse cascade of magnetic helicity driven by the chiral anomaly,''
  Phys.\ Rev.\ D {\bf 92}, no. 12, 125031 (2015)
  %doi:10.1103/PhysRevD.92.125031
  %[arXiv:1509.07790 [hep-th]].
  %%CITATION = doi:10.1103/PhysRevD.92.125031;%%
  %18 citations counted in INSPIRE as of 22 Nov 2016
  
  %\cite{Li:2016tel}
\bibitem{Li:2016tel} 
  H.~Li, X.~l.~Sheng and Q.~Wang,
  ``Electromagnetic fields with electric and chiral magnetic conductivities in heavy ion collisions,''
  Phys.\ Rev.\ C {\bf 94}, no. 4, 044903 (2016)
  %doi:10.1103/PhysRevC.94.044903
 % [arXiv:1602.02223 [nucl-th]].
  %%CITATION = doi:10.1103/PhysRevC.94.044903;%%
  %9 citations counted in INSPIRE as of 22 Nov 2016


%\cite{Tuchin:2016qww}
\bibitem{Tuchin:2016qww} 
  K.~Tuchin,
 ``Spontaneous topological transitions of electromagnetic fields in spatially inhomogeneous CP-odd domains,''  Phys.\ Rev.\ C {\bf 94}, no. 6, 064909 (2016)
 % doi:10.1103/PhysRevC.94.064909
  %[arXiv:1607.07481 [hep-ph]].
  %%CITATION = doi:10.1103/PhysRevC.94.064909;%%
  %1 citations counted in INSPIRE as of 07 Mar 2017  
  
  %\cite{Hirono:2016jps}
\bibitem{Hirono:2016jps} 
  Y.~Hirono, D.~E.~Kharzeev and Y.~Yin,
  ``Quantized chiral magnetic current from reconnections of magnetic flux,''
  Phys.\ Rev.\ Lett.\  {\bf 117}, no. 17, 172301 (2016)
  %doi:10.1103/PhysRevLett.117.172301
  %[arXiv:1606.09611 [hep-ph]].
  %%CITATION = doi:10.1103/PhysRevLett.117.172301;%%
  %1 citations counted in INSPIRE as of 22 Nov 2016



%\cite{Xia:2016any}
\bibitem{Xia:2016any} 
  X.~l.~Xia, H.~Qin and Q.~Wang,
  ``Approach to Chandrasekhar-Kendall-Woltjer State in a Chiral Plasma,''
  Phys.\ Rev.\ D {\bf 94}, no. 5, 054042 (2016)
  %doi:10.1103/PhysRevD.94.054042
  %[arXiv:1607.01126 [nucl-th]].
  %%CITATION = doi:10.1103/PhysRevD.94.054042;%%
  %1 citations counted in INSPIRE as of 22 Nov 2016

%\cite{Qiu:2016hzd}
\bibitem{Qiu:2016hzd} 
  Z.~Qiu, G.~Cao and X.~G.~Huang,
  ``On electrodynamics of chiral matter,''
  Phys.\ Rev.\ D {\bf 95}, no. 3, 036002 (2017)
  %doi:10.1103/PhysRevD.95.036002
  %[arXiv:1612.06364 [cond-mat.mes-hall]].
  %%CITATION = doi:10.1103/PhysRevD.95.036002;%%
  %1 citations counted in INSPIRE as of 01 Mar 2017

  
%%%%%  end EM fields and CME %%%%% 
  
  
 
%%%%%%%%  Begin Baryon Stopping %%%%%%%%
 
    %\cite{Kharzeev:1996sq}
\bibitem{Kharzeev:1996sq} 
  D.~Kharzeev,
  ``Can gluons trace baryon number?,''
  Phys.\ Lett.\ B {\bf 378}, 238 (1996)
 % doi:10.1016/0370-2693(96)00435-2
  %[nucl-th/9602027].
  %%CITATION = doi:10.1016/0370-2693(96)00435-2;%%
  %200 citations counted in INSPIRE as of 07 Sep 2017
  
  %\cite{Itakura:2003jp}
\bibitem{Itakura:2003jp} 
  K.~Itakura, Y.~V.~Kovchegov, L.~McLerran and D.~Teaney,
  ``Baryon stopping and valence quark distribution at small x,''
  Nucl.\ Phys.\ A {\bf 730}, 160 (2004)
  %doi:10.1016/j.nuclphysa.2003.10.016
  %[hep-ph/0305332].
  %%CITATION = doi:10.1016/j.nuclphysa.2003.10.016;%%
  %40 citations counted in INSPIRE as of 07 Sep 2017
  
%%%%%%%%  end Baryon Stopping %%%%%%%%


 

%%%%%% Begin BLAST WAVE MODEL %%%%%%%%%

%\cite{Siemens:1978pb}
\bibitem{Siemens:1978pb} 
  P.~J.~Siemens and J.~O.~Rasmussen,
  ``Evidence for a blast wave from compress nuclear matter,''
  Phys.\ Rev.\ Lett.\  {\bf 42}, 880 (1979).
  %doi:10.1103/PhysRevLett.42.880
  %%CITATION = doi:10.1103/PhysRevLett.42.880;%%
  %319 citations counted in INSPIRE as of 08 Sep 2017

%\cite{Teaney:2000cw}
\bibitem{Teaney:2000cw} 
  D.~Teaney, J.~Lauret and E.~V.~Shuryak,
  ``Flow at the SPS and RHIC as a quark gluon plasma signature,''
  Phys.\ Rev.\ Lett.\  {\bf 86}, 4783 (2001)
  %doi:10.1103/PhysRevLett.86.4783
  %[nucl-th/0011058].
  %%CITATION = doi:10.1103/PhysRevLett.86.4783;%%
  %605 citations counted in INSPIRE as of 08 Sep 2017

%\cite{Kolb:2000fha}
\bibitem{Kolb:2000fha} 
  P.~F.~Kolb, P.~Huovinen, U.~W.~Heinz and H.~Heiselberg,
  ``Elliptic flow at SPS and RHIC: From kinetic transport to hydrodynamics,''
  Phys.\ Lett.\ B {\bf 500}, 232 (2001)
  %doi:10.1016/S0370-2693(01)00079-X
 % [hep-ph/0012137].
  %%CITATION = doi:10.1016/S0370-2693(01)00079-X;%%
  %412 citations counted in INSPIRE as of 08 Sep 2017
  
  %%%%%% end BLAST WAVE MODEL %%%%%%%%%

  
  %\cite{Holliday:2016lbx}
\bibitem{Holliday:2016lbx} 
  R.~Holliday, R.~McCarty, B.~Peroutka and K.~Tuchin,
  ``Classical electromagnetic fields from quantum sources in heavy-ion collisions,''
  Nucl.\ Phys.\ A {\bf 957}, 406 (2017)
  %doi:10.1016/j.nuclphysa.2016.10.003
  %[arXiv:1604.04572 [hep-ph]].
  %%CITATION = doi:10.1016/j.nuclphysa.2016.10.003;%%
  %5 citations counted in INSPIRE as of 22 Nov 2016

%\cite{Peroutka:2017esw}
\bibitem{Peroutka:2017esw} 
  B.~Peroutka and K.~Tuchin,
  ``Quantum diffusion of electromagnetic fields of ultrarelativistic spin-half particles,''
  Nucl.\ Phys.\ A {\bf 966}, 64 (2017)
  %doi:10.1016/j.nuclphysa.2017.05.104
  %[arXiv:1703.02606 [hep-ph]].
  %%CITATION = doi:10.1016/j.nuclphysa.2017.05.104;%%
  %2 citations counted in INSPIRE as of 07 Sep 2017



%\cite{MF} 
\bibitem{MF}
P.~M.~Morse and H.~Feshbach, ``Methods of theoretical physics. Part I", McGraw-Hill (1953).

%\cite{Teaney:2003kp}
\bibitem{Teaney:2003kp} 
  D.~Teaney,
  ``The Effects of viscosity on spectra, elliptic flow, and HBT radii,''
  Phys.\ Rev.\ C {\bf 68}, 034913 (2003)
  %doi:10.1103/PhysRevC.68.034913
  %[nucl-th/0301099].
  %%CITATION = doi:10.1103/PhysRevC.68.034913;%%
  %651 citations counted in INSPIRE as of 11 Sep 2017


%\cite{Zakharov:2017gkb}
\bibitem{Zakharov:2017gkb} 
  B.~G.~Zakharov,
  ``Quantum analysis of fluctuations of electromagnetic fields in heavy-ion collisions,''
  JETP Lett.\  {\bf 105}, no. 12, 758 (2017)
  %doi:10.1134/S0021364017120037
  %[arXiv:1703.04271 [nucl-th]].
  %%CITATION = doi:10.1134/S0021364017120037;%%
  %1 citations counted in INSPIRE as of 08 Sep 2017

\end{thebibliography}
\end{document}